\documentclass[prl,twocolumn,showpacs,preprintnumbers,amsmath,amssymb]{revtex4}
\usepackage{amsmath}
\usepackage{graphicx}% Include figure files
\usepackage{dcolumn}% Align table columns on decimal point
\usepackage{bm}% bold math
\newcommand{\corr}[1]{#1}
\begin{document}

\title{The inverse cascade of magnetic helicity  in magnetohydrodynamic turbulence}
%\author{Wolf-Christian M\"uller, Shiva Kumar Malapaka\altaffilmark{1} \and Angela Busse\altaffilmark{2}}
\author{Wolf-Christian M\"uller}
\email{wolf.mueller@ipp.mpg.de}
\affiliation{Max-Planck-Institut f\"ur Plasmaphysik, 85748 Garching, Germany}
\author{Shiva Kumar Malapaka}
\email{shiva.kumar.malapaka@ipp.mpg.de}
\altaffiliation[Now at:]{University of Leeds, Faculty of Mathematics and Physical Sciences, United\affiliation{Max-Planck-Institut f\"ur Plasmaphysik, 85748 Garching, Germany}
 Kingdom}
\author{Angela Busse}
\email{a.busse@soton.ac.uk}
\altaffiliation[Now at:]{University of Southampton, Faculty of Engineering and the Environment, United Kingdom}
\affiliation{Max-Planck-Institut f\"ur Plasmaphysik, 85748 Garching, Germany}
\begin{abstract}
The nonlinear dynamics of magnetic helicity, $H^M$, 
which is responsible for large-scale magnetic structure formation in electrically conducting 
turbulent media is investigated 
in forced and decaying 
three-dimensional magnetohydrodynamic turbulence. This is done
with the help of  high resolution direct numerical simulations and statistical closure theory.
The numerically observed spectral scaling of $H^M$ is at variance with 
earlier work using a statistical closure model [Pouquet et al., J. Fluid Mech. \textbf{77} 321 (1976)]. 
By revisiting this theory a universal dynamical balance relation is found
that includes effects of kinetic helicity, as well as kinetic and magnetic energy on the 
inverse cascade of $H^M$ and explains the above-mentioned discrepancy. 
%The reason for the mentioned discrepancy is identified 
%in the previously assumed equipartition of kinetic and magnetic energy and the additional
%influence of kinetic helicity. 
Considering the result
in the context of mean-field dynamo theory suggests a nonlinear modification of the $\alpha$-dynamo effect
important in the context of magnetic field excitation in turbulent plasmas.
\end{abstract}

\pacs{47.27.-i, 52.65.Kj, 47.27.Gs, 47.65.Md}
\maketitle

The emergence of large-scale magnetic structures in turbulent plasmas
is dynamically important in many astrophysical settings, e.g. with
regard to the interstellar medium 
or the magnetic field generation in planets and stars by the 
turbulent dynamo effect, e.g. \cite{soward_weiss:dynreview,brandenburg:dynamoreview,kulsrud_zweibel:cosmicdynreview}.
The structure formation can be studied via the magnetic helicity 
$H^\text{M} =\frac{1}{2V}\int_{V}{dV\bf{a}\cdot\bf{b}}$ where $\bf{b}$ is the
magnetic field and $\bf{a}$ denotes the magnetic vector potential.
This topological characteristic of magnetic fields yields a measure of 
the linkage and the twist of the field lines, \citep{moffat:maghel,berger_field:maghel}.
In the magnetohydrodynamic (MHD) single-fluid
approximation \citep{biskamp:book} which neglects microscopic scales and
the associated kinetic dynamics $H^\text{M}$ is ideally conserved in a three-dimensional volume 
with 
periodic or closed boundary conditions 
\citep{woltjer:invariants}. It is thus  
prone to a nonlinear and conservative inverse spectral cascade process in the inertial range of 
MHD plasma turbulence. 
If driven at small scales, $\ell$, the cascade results in spectral transfer 
of magnetic helicity towards small spatial wavenumbers $k\sim \ell^{-1}$ \citep{pouquet_frisch:invcasceq}, 
i.e.
to the formation of large-scale magnetic structures. 
This process is thus %\corr{be} 
of fundamental importance with regard to, e.g., the dynamics of 
magnetic fields in the above-mentioned turbulent astrophysical settings.
In spite of its importance little is known about the nonlinear dynamics which underlies
the inverse cascade. It is the \corr{purpose} of this work to shed some light on the rather mysterious 
nonlinear phenomenon creating large-scale order out of quasi-random turbulent magnetic 
fluctuations. Please note that the constraining effect of magnetic helicity conservation on 
certain $\alpha$-dynamo configurations is,
although important, 
beyond the scope of this work, cf., for example, \cite{brandenburg:dyntheoreview}.    
The existence of such an inverse cascade was first
demonstrated in numerical simulations based on the eddy damped quasi
normal Markovian (EDQNM) closure model of three-dimensional MHD turbulence
\citep{pouquet_frisch:invcasc} which is to the best of our knowledge the only work 
dealing theoretically with the spectral self-similarity of magnetic helicity. 
The associated 
self-similar spectral signature in the turbulent inertial range, $\sim k^{-2}$, is 
in agreement with dimensional analysis based on a constant nonlinear spectral flux.  
Several studies applying direct numerical simulations (DNS) find
inverse transfer of magnetic helicity, see e.g. 
\cite{brandenburg:invcascdynamo,alexakis_mininni_pouquet:invcasc, mininni_pouquet:inverse},
without reporting self-similar scaling behaviour, a notable exception being \cite{mininni_pouquet:inverse}. 

In this Rapid Communication the inverse cascade of magnetic helicity in homogeneous MHD turbulence
is studied by three-dimensional high-resolution direct numerical simulations. 
In the main setup kinetic and magnetic energy and magnetic helicity are injected at small 
scales of the initially excited spectral range of turbulent fluctuations.
%Consequently, in case (a) the inverse cascade of
%$H^\mathrm{M}$ proceeds on scales larger than the direct turbulent
%energy cascade while in configuration (b) featuring decaying
%turbulence the inertial scaling ranges of both cascades overlap.
%The observed self-similar scaling is not in agreement with 
%the above-mentioned literature. 
It is shown that macroscopic quantities, in particular kinetic 
helicity and the ratio of kinetic and magnetic energy,
have an important influence on the $H^\mathrm{M}$-cascade that is captured by a 
universal relation based on dimensional analysis of the MHD-EDQNM closure model.
This insight suggests a new link between 
magnetic helicity and mean-field dynamo theory in particular with regard to the 
saturation behaviour of the dynamo mechanism. 

The dimensionless MHD equations are written as
\begin{equation}\label{1}
\partial_t\boldsymbol{\omega} = \nabla\times({\bf v}\times\boldsymbol{\omega} + 
{\bf j}\times{\bf b})+ ({\mu}_n(-1)^{n-1}\Delta^n +\corr{\lambda}\Delta^{-2})\boldsymbol{\omega}+{\bf F_{v}}\,,
\end{equation}
\begin{equation}\label{2}
 \partial_{t}{\bf b} = \nabla\times({\bf v} \times {\bf b}) + ({\eta}_n(-1)^{n-1}\Delta^n +\corr{\lambda}\Delta^{-2}){\bf b}+{\bf F_{b}}\,, 
 \end{equation}
 \begin{equation}\label{3}
\nabla\cdot\bf{v} = \nabla\cdot\bf{b}= 0,
\end{equation} 
where $\bf{v}$ is the velocity, $\boldsymbol{\omega}=\nabla \times \bf{v}$ the vorticity,
$\bf{b}$ the magnetic field, and $\bf{j}=\nabla \times \bf{b}$ the electric current density.
Equations (\ref{1})--(\ref{3}) are solved by a standard
pseudospectral method using a leapfrog scheme for
time integration. 
Anti-aliasing is achieved by spherical mode truncation. The simulation domain is a triply $2\pi$-periodic cube discretized by $1024^3$ collocation
points.
Hyperviscous small-scale dissipation operators of order $n=8$ 
are used to improve scale separation parametrized by  
the hyperdiffusion coefficients ${\mu}_n$ and ${\eta}_n$ with the hyperviscosity ${\mu_8}=9\cdot10^{-41}$ and 
${\mu}_8/{\eta}_8=1$.
Boundary effects at smallest spatial wavenumbers $k$ are alleviated by a large-scale energy sink
$\corr{\lambda}\Delta^{-2}$ 
for both fields with the constant $\corr{\lambda}$ 
set to $0.5$. The forcing terms ${\bf F_{v}}$ and
${\bf F_{b}}$ are random, delta-correlated processes of equal amplitude that act
over a band of wave numbers $k\in[k_0-3,k_0+3]$ with $k_0=206$.
%,$70$(b). 
They inject velocity- and magnetic-field fluctuations with well defined \corr{magnetic and 
kinetic} helicity, kinetic helicity being defined as $H^{K} = \frac{1}{2V}\int_{\it{V}}{\it{dV}\bf{v}\cdot\bf{\boldsymbol{\omega}}}$.
Such driving, chosen here for simplicity and numerical efficiency, 
could in principle be realized by a random small-scale distribution of electric currents
and forces.
The initial velocity and magnetic fields are smooth with
equal energies, random phases and fluctuations that have a Gaussian energy distribution,
peaked at $k_0$. In the course of the simulation 
the total energy quickly attains a 
quasi-stationary state, fluctuating around unity with $E^M/E^K\approx 9$.
% in case (b)
%total energy decays self-similarly from unity to about 0.04 over the simulation period of seven %large-eddy turnover 
%times where $E^\mathrm{M}/E^\mathrm{K}$ stays roughly constant at about $xx$.
Cross-helicity, $H^{C} =\frac{{1}}{{2V}}\int_{{V}}{{dV}\bf{v}\cdot\bf{b}}$, is
negligible.
% in both cases. 
The simulation is carried up to t=6.66 large-eddy turnover times.
%while decaying turbulence simulations are carried up to t=10. 
The application of hyperviscous dissipation operators while necessary to observe well-developed
scaling ranges precludes the unambiguous definition of a Reynolds number.

\begin{figure}
\includegraphics[width=8cm]{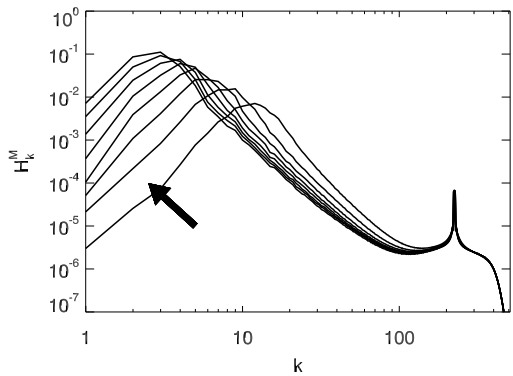}
%\includegraphics{f1.eps}
%a) \includegraphics[width=6cm,viewport=150 133 450 545,clip]{f11.pdf}
%b)\includegraphics[width=6cm,viewport=0 0 300 300,clip]{f12.eps}
\caption{\label{fig:f1}Inverse cascade of magnetic helicity, $H^\text{M}_k$, 
in 3D-MHD turbulence for $t\in [1,6.66]$. Curves are given for points in time spread equidistantly over the interval.
Magnetic fluctuations with maximal magnetic helicity and non-helical velocity fluctuations 
are supplied through the forcing with $k_0=206$.
}
\end{figure}

\begin{figure}
 \includegraphics[width=8cm,height=7cm]{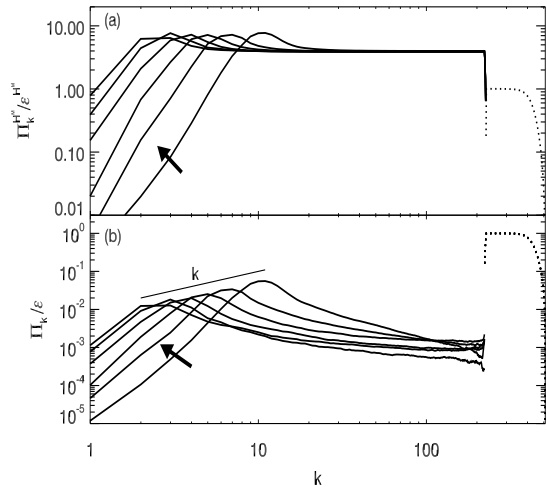}
\caption{\label{fig:f2} Moduli of fluxes of magnetic helicity (a) and total energy (b) (inverse: solid curves, direct: dotted curves) normalized by corresponding dissipation rates. 
Flux spectra are shown for $t\in [1.66,6.66]$ for the forced simulation and are spread equidistantly over this period.}
\end{figure}

The temporal evolution of the 
magnetic helicity spectrum over the simulation period shown in Fig. \ref{fig:f1} 
indicates inverse spectral transfer. 
%In the decaying
%case while the initial high energy state at moderate scales start
%decaying, the magnetic helicity moves towards the large scales.
This is also reflected by the nonlinear spectral flux of magnetic helicity, 
$\Pi^{{H^{M}}}_k =
2\int^{{k}}_{0}dk'k'^2\int d\Omega({\bf{b}}^{*}_{\mathbf{k}'}\cdot[\mathbf{v}\times\mathbf{b}]_{\mathbf{k}'}+\text{c.c.})$ with 
`$[\bullet]_\mathbf{k}$' denoting Fourier transformation and `$*$' standing for complex conjugate ($\text{c.c.}$). 
The flux spectrum shown in Fig. \ref{fig:f2}(a),  is constant over finite wavenumber intervals on both sides of the 
forcing band signalling equilibrium between turbulence driving and dissipation. 
While on the right-hand-side direct spectral transfer is observed as a result of 
the small-scale energy sink, on the left-hand-side of the forcing band 
an inverse cascade develops which is driven by the 
constant magnetic helicity input around $k_0$.  
The spectral flux of total energy, 
$E=E^K+E^M=\int d^3k(|v_\mathbf{k}|^2+|b_\mathbf{k}|^2)/2$, is given by 
$\Pi_k =
\int^{{k}}_{0}dk'k'^2\int d\Omega(%i/k'^2
i\mathbf{k}'\times[\mathbf{v\times\boldsymbol{\omega}+\mathbf{j}\times\mathbf{b}}]_{\mathbf{k}'}\cdot\mathbf{v}^*_{\mathbf{k}'}
+i\mathbf{k}'\times[\mathbf{v}\times\mathbf{b}]_{\mathbf{k}'}\cdot\mathbf{b}^*_{\mathbf{k}'}+ \text{c.c.})$ and is shown in 
Fig. \ref{fig:f2}(b). It lacks spectral constancy in the inverse cascade region and  
it is principally carried by  magnetic energy transfer. This suggests that the inverse energy 
flux is  a consequence of the inverse cascade of $H^\mathrm{M}$. 
The linear scaling of the $\Pi_k$-envelope which follows from the dimensional 
estimate $E_k\sim k H^\mathrm{M}_k$
in combination with the approximate constant value of the envelope of $\Pi_k^{H^\mathrm{M}}$ 
also supports this interpretation. 
  
\begin{figure}
 \includegraphics[width=8cm,height=8cm]{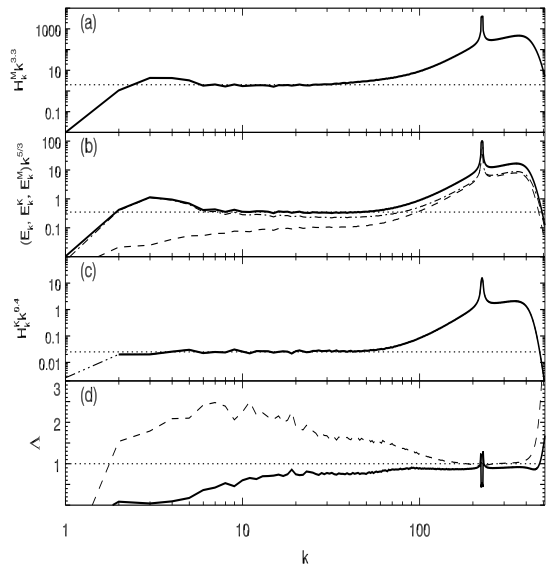}
\caption{\label{fig:f3} Compensated spectra of (a) magnetic helicity , (b) total (solid line), kinetic (dashed line), magnetic (dashed-dotted line) energy, (c) kinetic helicity. (d) $\Lambda=(E^K_k/E^M_k)^\gamma H^J_k/H^K_k$ for $\gamma=1$ (dashed line) and $\gamma=2$ at t=6.66 in the driven simulation.}
\end{figure}

The compensated  magnetic helicity spectrum at the end of the simulation period is displayed in Fig. 
\ref{fig:f3}(a).
It exhibits two approximate
scaling ranges: on the direct transfer side for $250\lesssim k\lesssim 400$ and in the
inverse transfer region, $7 \lesssim k\lesssim 30$. 
The corresponding asymptotic scaling laws are $H_k^\mathrm{M}\sim k^{-3.3}$ (inverse, cf. 
\cite{mininni_pouquet:inverse}) and $H_k^\mathrm{M}\sim k^{-1.5}$ (direct).
The latter value has to be taken with care due to the very short spectral range and the high-order hyperviscosity 
acting at largest wavenumbers.
The inverse cascade scaling is at variance with the $k^{-2}$-behaviour reported in
\cite{pouquet_frisch:invcasc}.
 
Since $H^\mathrm{M}$ is the helicity of the magnetic vector potential, $\mathbf{a}$, 
its inverse cascade pulls quantities derived from this potential, 
e.g. magnetic energy and to a lesser extent also
electric current density, $\mathbf{j}_k\sim k^2\mathbf{a}_k$, \corr{towards} large scales.
Magnetic and velocity field are intrinsically coupled in MHD turbulence by Alfv\'enic 
fluctuations 
\citep{mueller_grappin:endyn} and thus similar behaviour is observed for the spectral kinetic energy and 
kinetic helicity, $H^\mathrm{K}_k$, as well.
The respective spectra are thus expected to inherit the self-similarity 
from $H_k^\mathrm{M}$ as is indeed observed, see, e.g., Fig. \ref{fig:f3}(c).
It seems to be reasonable to regard the merging of current-carrying structures  
where the currents have significant 
positively aligned components and thus experience mutual attraction as the  physical mechanism 
of the inverse cascade of magnetic helicity, cf. \cite{biskamp_bremer:invcascade}. 
This is also in line with recent statements in the literature about non-locality of 
the magnetic helicity flux
\cite{alexakis_mininni_pouquet:invcasc,aluie_eyink:MHDlocality}.%,rammah_malapaka_mueller:MHDlocality}.
Terming the inverse transfer of $H^\text{M}_k$ a ``cascade'' is thus merely
following convention and not a physical description of the actual process of a spectrally non-local merging 
of current-carrying strucures. 
Note that in this simulation no kinetic helicity is injected by the turbulence driving. 
In the case of simultaneous injection of kinetic helicity the spectral diagnostics discussed in 
this work show no significant difference to the presented simulations.
In general the details of the small-scale forcing, like randomness of amplitude and/or complex
phases, were verified to have no measureable impact on the self-similar behavior reported below.
The only significant parameter in this respect 
is the separation of the smallest admissible wavenumber, $k=1$, and the forcing-wavenumber of 
about two decades. 
The scale-separation
determines the extent of the observable self-similar inverse-cascade dynamics. 
It is thus even more important than the classical Reynolds number which is not well defined 
due to the necessary 
high-order hyperviscous small-scale diffusion.    
 
The energy spectra in Fig. \ref{fig:f3}(b) exhibit approximate scaling known from decaying turbulence, cf., e.g., 
\cite{mueller_biskamp:3dmhdscale,mueller_grappin:endyn}, i.e. $E_k\sim k^{-5/3}$ and 
$E^R_k=|E_k^M-E_k^K|\sim k^{-7/3}$ (not shown) with an 
excess of magnetic energy.
The observations are in agreement with 
the interpretation of the finite levels of $E^K_k$ and of $H^K_k$ as a result of 
the local and temporary stirring induced by changes of magnetic-field topology. This is to
be expected in the course of the inverse cascade of magnetic helicity. 
As will be shown in the following, the lacking equipartition 
of $E_k^K$ and $E_k^M$ and the presence of kinetic helicity  
are the reasons of disagreement with the above-mentioned EDQNM simulations of \cite{pouquet_frisch:invcasc}. There, the relaxation time of nonlinear interaction, $\theta_{kpq}$, which represents a free parameter 
of the EDQNM approach and determines the nonlinear process governing turbulent dynamics, 
is chosen to be the Alfv\'en time, $(kB_0)^{-1}$. Consequently, 
the resulting dominance of Alfv\'enic 
interactions drives the system in the inertial range into nearly perfect equipartition
of kinetic and magnetic \corr{energies}.

The observed spectral scaling of magnetic helicity can be better understood with the help of 
the integro-differential EDQNM equation for the evolution of $H^M_k$.
A formally similar approach has been successful with regard to the residual energy spectrum 
$E_k^R=|E^M_k-E^K_k|$ \citep{mueller_grappin:endyn}.
The equation for the evolution of $H_k^M$ in the EDQNM model \citep{pouquet_frisch:invcasc}
reads
\begin{equation}
(\partial_t+\eta_1k^2)H_k^M=\int_\triangle dpdq\theta_{kpq}(T^\mathrm{adv}_k+T^\mathrm{khl}_k+T^\mathrm{lor}_k)\label{edqnmmh}
\end{equation} with
\begin{eqnarray*}
T^\mathrm{adv}_k&=h_{kpq}\frac{k}{pq}(k^2E^K_qH^M_p-p^2E^K_qH^M_k)\,,\\
T^\mathrm{khl}_k&=h_{kpq}\frac{k}{pq}(\frac{k^2}{p^2}E^M_qH^K_p-\frac{k^2}{p^2}E^M_kH^K_q)\,,\\
T^\mathrm{lor}_k&=e_{kpq}\frac{p^2}{k}E^M_kH^M_q-j_{kpq}\frac{kp}{q}E^M_qH^M_k\,.
\end{eqnarray*}
The geometric coefficients $h_{kpq}$, $e_{kpq}$, and $j_{kpq}$
follow from the solenoidality constraints (\ref{3}) and are given  
in \cite{pouquet_frisch:invcasc}. The \lq$\triangle$' restricts integration
to {wave vectors} $\textbf{k}$, $\textbf{p}$, $\textbf{q}$ which form a triangle, i.e. to a domain
in the $p$-$q$ plane which is defined by $q=|\textbf{p}+\textbf{k}|$.
The time $\theta_{kpq}$ is characteristic of the eddy damping of the nonlinear
energy flux involving wave numbers $k$, $p$, and $q$. 
It is defined phenomenologically but its particular form
does not play a role in the following arguments. 

The three nonlinear contributions on the right-hand-side of Eq. (\ref{edqnmmh})
can be associated with the advective ($T^\mathrm{adv}_k$) and explicitly twisting 
($T_k^\mathrm{khl}$) effects of turbulent fluctuations, as well as self-interaction  
($T^\mathrm{Lor}_k$) of the magnetic field through the Lorentz force.
Assuming that the most important nonlinearities involve the turbulent velocity and that 
the spectral scaling range of $H_k^M$ is stationary, a dynamical 
equilibrium of  turbulent advection and the $H^M$-increasing effect
of helical fluctuations is assumed, \corr{i.e.} $T_k^\mathrm{adv}\sim T_k^\mathrm{khl}$.
Dimensional approximation of the respective flux terms, $kE_k^KH^M_k\sim k^{-1}E^M_kH^K_k$, 
yields
\begin{equation}
H^K_k\sim \left(\frac{E^K_k}{E^M_k}\right)^\gamma k^2 H_k^M\,,\quad\gamma=1\,. \label{mainres}
\end{equation} 
This is a statement about the spectral dynamics of kinetic and magnetic \corr{helicities} 
(or \corr{kinetic and} current \corr{helicities}, $H^J=\frac{1}{2V}\int_V(\nabla\times\mathbf{b})\cdot\mathbf{b}$, 
since $H^J_k\sim k^2H^M_k$) in the case of $E_k^K/E_k^M\neq 1$.
The agreement of Rel. (\ref{mainres}) with the numerical experiment is 
significantly improved, see Fig. \ref{fig:f3}(d), when increasing $\gamma$ by one. This   
yields the main result     
\begin{equation} 
H^K_k\sim \left(\frac{E^K_k}{E^M_k}\right)^2H_k^J\,.\label{main2res}
\end{equation}
Rel. (\ref{main2res}) is fulfilled, i.e. constant and close to unity, for almost all wavenumbers $k>12$ excluding the drive and deep dissipation scales. It does however not cover the full spectral scaling range of $H^M_k$ 
due to its susceptibility to the asymmetry of energies and helicities introduced by the 
large-scale energy sink.
The higher-order modification of Rel. (\ref{mainres}) can not
be motivated within the framework of quasi-normal EDQNM theory using the chosen approach of 
nonlinear equilibrium. The underlying cause is presumably the  non-locality of the 
inverse cascade process which is not captured by the dimensional simplification of the EDQNM
equations. 

For verification purposes a test simulation of decaying turbulence with the same numerical resolution of $1024^3$ and 
an initially  finite level of magnetic
helicity, i.e. 50\% of the energetically possible maximum  $H^M_\text{max}\sim E^M/k_0$, is conducted. 
The initial Gaussian random-phase energy spectrum with $E^K_k=E^M_k$ for all $k$ is peaked around $k_0=70$ to allow some 
development of inverse transfer.
The distribution of $H^M_k$ is homogeneous over the initial spectrum. The hyperdiffusive coefficients are
chosen as $\mu_8=\eta_8=3\cdot10^{-41}$ with forcing switched off.
\begin{figure}
\includegraphics[width=8cm,height=7cm]{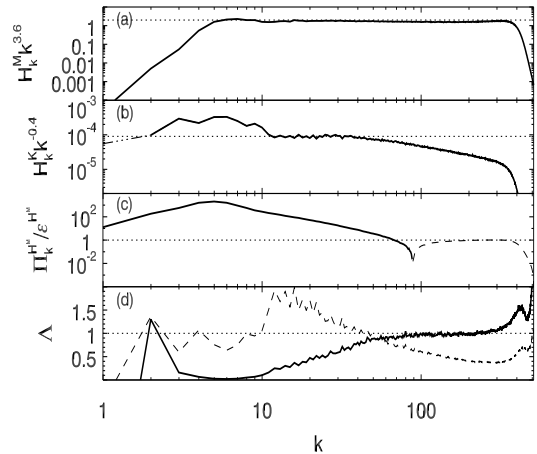}
\caption{\label{fig:f5} Compensated spectra of (a) magnetic helicity , (b) kinetic helicity. (c) Spectral magnetic helicity flux normalized by corresponding dissipation rate (Inverse: solid curve, direct: dashed curve). (d) $\Lambda=(E^K_k/E^M_k)^\gamma H^J_k/H^K_k$ for $\gamma=1$ and scaled by a factor $1/3$ (dashed) and for $\gamma=2$. All spectra shown have been taken at $t=9.2$ in the decaying simulation.}
\end{figure}    
Fig. \ref{fig:f5} displays the most important results of this decaying test simulation after 
about 9.2 large-eddy turnover times. The magnetic helicity, Fig. \ref{fig:f5}(a), exhibits approximate self-similar 
spectral scaling $\sim k^{-3.6}$ with most of the excited scales in the  
range $15\lesssim k\lesssim 60$ 
having developed during the decay. Kinetic helicity, see Fig. \ref{fig:f5}(b), reflects this as it is 
not injected into the system but generated via the Lorentz force during the inverse transfer of
$H_k^M$. It shows two distinct asymptotic power-laws: $\sim k^{-0.5}$ for $60\lesssim k \lesssim 120$
and $\sim k^{0.4}$ for $11\lesssim k \lesssim 45$ in the range
covered by the inverse transfer of $H_k^M$ during turbulence decay.
\corr{The signatures of kinetic, magnetic and total energies are} similar to the observation made in 
the driven case. The spectral flux of magnetic helicity, Fig. \ref{fig:f5}(c), exhibits a split 
near $k_0$ as for $k>k_0$ small-scale dissipation determines the helicity transfer direction while for $k<k_0$
an inverse transfer with $k$-dependent $\Pi_k^{H^M}$ is observed.
This is reflected by Rel. (\ref{main2res}), cf. Fig. \ref{fig:f5}(d), which is only fulfilled in the region of 
approximately constant $\Pi_k^{H^M}$ and is apparently  independent of the transfer direction.  

Comparison of the presented findings with mean-field dynamo theory, cf., for example, \cite{brandenburg:dynamoreview} and references therein, in particular
the $\alpha$-coefficient including the current helicity, $\alpha\sim H^J-H^K$, is interesting. 
Rel. (\ref{main2res}) is consistent with vanishing $\alpha$ since the cascade process does not 
generate magnetic flux and, thus, does not act as a dynamo itself. It furthermore suggests a 
modified $\alpha\sim (E^K/E^M)^2H^J-H^K$ which includes the squared ratio of kinetic and magnetic energy $\sim (E^M)^{-2}$
as a turbulent dynamo-quenching mechanism of the current helicity contribution to $\alpha$. 
This form of $\alpha$-quenching has recently been observed in  a numerical test-field model
\cite{rheinhardt_brandenburg:alphaquenchb-4}.  
  
In summary a new and probably universal relation connecting the
spectral behaviour of magnetic and kinetic helicities and energies in
homogeneous MHD turbulence is found motivated by statistical EDQNM
closure results.  DNS of MHD turbulence that is decaying or driven at
small scales confirm the validity of the findings for spectral
intervals of constant flux of magnetic helicity.  The result has
interesting connections to the $\alpha$ coefficient known from
mean-field dynamo theory: it suggests an inherent and strong quenching of the dynamo,
in particular of the current helicity effect, if the energy of turbulent magnetic 
fluctuations grows compared to the kinetic contribution of velocity.

\begin{acknowledgments}
%MSK wants to thank A.Busse for her support.
\corr{The authors are grateful for discussions with A. Brandenburg, P. Diamond, U. Frisch, R. Grappin, 
P. Mininni and A. Pouquet.}
\end{acknowledgments}
%\bibliography{/home/Wolf/Texte/Bibliographien/Dynamo,/home/Wolf/Texte/Bibliographien/Turbulence}

\begin{thebibliography}{18}
\expandafter\ifx\csname natexlab\endcsname\relax\def\natexlab#1{#1}\fi
\expandafter\ifx\csname bibnamefont\endcsname\relax
  \def\bibnamefont#1{#1}\fi
\expandafter\ifx\csname bibfnamefont\endcsname\relax
  \def\bibfnamefont#1{#1}\fi
\expandafter\ifx\csname citenamefont\endcsname\relax
  \def\citenamefont#1{#1}\fi
\expandafter\ifx\csname url\endcsname\relax
  \def\url#1{\texttt{#1}}\fi
\expandafter\ifx\csname urlprefix\endcsname\relax\def\urlprefix{URL }\fi
\providecommand{\bibinfo}[2]{#2}
\providecommand{\eprint}[2][]{\url{#2}}

\bibitem[{\citenamefont{Soward et~al.}(2005)\citenamefont{Soward, Jones,
  Hughes, and Weiss}}]{soward_weiss:dynreview}
\bibinfo{editor}{\bibfnamefont{A.~M.} \bibnamefont{Soward}},
  \bibinfo{editor}{\bibfnamefont{C.~A.} \bibnamefont{Jones}},
  \bibinfo{editor}{\bibfnamefont{D.~W.} \bibnamefont{Hughes}},
  \bibnamefont{and} \bibinfo{editor}{\bibfnamefont{N.~O.} \bibnamefont{Weiss}},
  eds., \emph{\bibinfo{title}{Fluid Dynamics and Dynamos in Astrophysics and
  Geophysics}}, Durham Symposium on Astrophysical Fluid Mechanics 2002
  (\bibinfo{publisher}{CRC Press}, \bibinfo{address}{Boca Raton, Florida},
  \bibinfo{year}{2005}).

\bibitem[{\citenamefont{Kulsrud and
  Zweibel}(2008)}]{kulsrud_zweibel:cosmicdynreview}
\bibinfo{author}{\bibfnamefont{R.~M.} \bibnamefont{Kulsrud}} \bibnamefont{and}
  \bibinfo{author}{\bibfnamefont{E.~G.} \bibnamefont{Zweibel}},
  \bibinfo{journal}{Reports on Progress in Physics}
  \textbf{\bibinfo{volume}{71}}, \bibinfo{pages}{046901}
  (\bibinfo{year}{2008}).

\bibitem[{\citenamefont{Brandenburg and
  Subramanian}(2005)}]{brandenburg:dynamoreview}
\bibinfo{author}{\bibfnamefont{A.}~\bibnamefont{Brandenburg}} \bibnamefont{and}
  \bibinfo{author}{\bibfnamefont{K.}~\bibnamefont{Subramanian}},
  \bibinfo{journal}{Physics Reports} \textbf{\bibinfo{volume}{417}},
  \bibinfo{pages}{1} (\bibinfo{year}{2005}).

\bibitem[{\citenamefont{Moffatt}(1969)}]{moffat:maghel}
\bibinfo{author}{\bibfnamefont{H.~K.} \bibnamefont{Moffatt}},
  \bibinfo{journal}{Journal of Fluid Mechanics} \textbf{\bibinfo{volume}{35}},
  \bibinfo{pages}{117} (\bibinfo{year}{1969}).

\bibitem[{\citenamefont{Berger and Field}(1984)}]{berger_field:maghel}
\bibinfo{author}{\bibfnamefont{M.~A.} \bibnamefont{Berger}} \bibnamefont{and}
  \bibinfo{author}{\bibfnamefont{G.~B.} \bibnamefont{Field}},
  \bibinfo{journal}{Journal of Fluid Mechanics} \textbf{\bibinfo{volume}{147}},
  \bibinfo{pages}{133} (\bibinfo{year}{1984}).

\bibitem[{\citenamefont{Biskamp}(1993)}]{biskamp:book}
\bibinfo{author}{\bibfnamefont{D.}~\bibnamefont{Biskamp}},
  \emph{\bibinfo{title}{Nonlinear Magnetohydrodynamics}}
  (\bibinfo{publisher}{Cambridge University Press},
  \bibinfo{address}{Cambridge}, \bibinfo{year}{1993}).

\bibitem[{\citenamefont{Woltjer}(1958)}]{woltjer:invariants}
\bibinfo{author}{\bibfnamefont{L.}~\bibnamefont{Woltjer}},
  \bibinfo{journal}{Proceedings of the National Academy of Sciences}
  \textbf{\bibinfo{volume}{44}}, \bibinfo{pages}{833} (\bibinfo{year}{1958}).

\bibitem[{\citenamefont{Frisch et~al.}(1975)\citenamefont{Frisch, Pouquet,
  L{\'e}orat, and Mazure}}]{pouquet_frisch:invcasceq}
\bibinfo{author}{\bibfnamefont{U.}~\bibnamefont{Frisch}},
  \bibinfo{author}{\bibfnamefont{A.}~\bibnamefont{Pouquet}},
  \bibinfo{author}{\bibfnamefont{J.}~\bibnamefont{L{\'e}orat}},
  \bibnamefont{and} \bibinfo{author}{\bibfnamefont{A.}~\bibnamefont{Mazure}},
  \bibinfo{journal}{Journal of Fluid Mechanics} \textbf{\bibinfo{volume}{68}},
  \bibinfo{pages}{769} (\bibinfo{year}{1975}).

\bibitem[{\citenamefont{Brandenburg}(2009)}]{brandenburg:dyntheoreview}
\bibinfo{author}{\bibfnamefont{A.}~\bibnamefont{Brandenburg}},
  \bibinfo{journal}{Space Science Reviews} \textbf{\bibinfo{volume}{144}},
  \bibinfo{pages}{87} (\bibinfo{year}{2009}).

\bibitem[{\citenamefont{Pouquet et~al.}(1976)\citenamefont{Pouquet, Frisch, and
  L\'eorat}}]{pouquet_frisch:invcasc}
\bibinfo{author}{\bibfnamefont{A.}~\bibnamefont{Pouquet}},
  \bibinfo{author}{\bibfnamefont{U.}~\bibnamefont{Frisch}}, \bibnamefont{and}
  \bibinfo{author}{\bibfnamefont{J.}~\bibnamefont{L\'eorat}},
  \bibinfo{journal}{Journal of Fluid Mechanics} \textbf{\bibinfo{volume}{77}},
  \bibinfo{pages}{321} (\bibinfo{year}{1976}).

\bibitem[{\citenamefont{Brandenburg}(2001)}]{brandenburg:invcascdynamo}
\bibinfo{author}{\bibfnamefont{A.}~\bibnamefont{Brandenburg}},
  \bibinfo{journal}{The Astrophysical Journal} \textbf{\bibinfo{volume}{550}},
  \bibinfo{pages}{824} (\bibinfo{year}{2001}).

\bibitem[{\citenamefont{Alexakis et~al.}(2006)\citenamefont{Alexakis, Mininni,
  and Pouquet}}]{alexakis_mininni_pouquet:invcasc}
\bibinfo{author}{\bibfnamefont{A.}~\bibnamefont{Alexakis}},
  \bibinfo{author}{\bibfnamefont{P.}~\bibnamefont{Mininni}}, \bibnamefont{and}
  \bibinfo{author}{\bibfnamefont{A.}~\bibnamefont{Pouquet}},
  \bibinfo{journal}{The Astrophysical Journal} \textbf{\bibinfo{volume}{640}},
  \bibinfo{pages}{335} (\bibinfo{year}{2006}).

\bibitem[{\citenamefont{Mininni and Pouquet}(2009)}]{mininni_pouquet:inverse}
\bibinfo{author}{\bibfnamefont{P.}~\bibnamefont{Mininni}} \bibnamefont{and}
  \bibinfo{author}{\bibfnamefont{A.}~\bibnamefont{Pouquet}},
  \bibinfo{journal}{Physical Review E} \textbf{\bibinfo{volume}{80}},
  \bibinfo{pages}{025401} (\bibinfo{year}{2009}).

\bibitem[{\citenamefont{M{\"u}ller and Grappin}(2005)}]{mueller_grappin:endyn}
\bibinfo{author}{\bibfnamefont{W.-C.} \bibnamefont{M{\"u}ller}}
  \bibnamefont{and} \bibinfo{author}{\bibfnamefont{R.}~\bibnamefont{Grappin}},
  \bibinfo{journal}{Physical Review Letters} \textbf{\bibinfo{volume}{95}},
  \bibinfo{pages}{114502} (\bibinfo{year}{2005}).

\bibitem[{\citenamefont{Biskamp and 
  Bremer}(1993)}]{biskamp_bremer:invcascade}
\bibinfo{author}{\bibfnamefont{D.}~\bibnamefont{Biskamp{ \nop{a}und U.
  Bremer}}}, \bibinfo{journal}{Physical Review Letters}
  \textbf{\bibinfo{volume}{72}}, \bibinfo{pages}{3819} (\bibinfo{year}{1993}).

\bibitem[{\citenamefont{Aluie and Eyink}(2010)}]{aluie_eyink:MHDlocality}
\bibinfo{author}{\bibfnamefont{H.}~\bibnamefont{Aluie}} \bibnamefont{and}
  \bibinfo{author}{\bibfnamefont{G.~L.} \bibnamefont{Eyink}},
  \bibinfo{journal}{Physical Review Letters} \textbf{\bibinfo{volume}{104}},
  \bibinfo{pages}{081101} (\bibinfo{year}{2010}).

\bibitem[{\citenamefont{M{\"u}ller and
  Biskamp}(2000)}]{mueller_biskamp:3dmhdscale}
\bibinfo{author}{\bibfnamefont{W.-C.} \bibnamefont{M{\"u}ller}}
  \bibnamefont{and} \bibinfo{author}{\bibfnamefont{D.}~\bibnamefont{Biskamp}},
  \bibinfo{journal}{Physical Review Letters} \textbf{\bibinfo{volume}{84}},
  \bibinfo{pages}{475} (\bibinfo{year}{2000}).

\bibitem[{\citenamefont{Rheinhardt and
  Brandenburg}(2010)}]{rheinhardt_brandenburg:alphaquenchb-4}
\bibinfo{author}{\bibfnamefont{M.}~\bibnamefont{Rheinhardt}} \bibnamefont{and}
  \bibinfo{author}{\bibfnamefont{A.}~\bibnamefont{Brandenburg}},
  \bibinfo{journal}{Astronomy \& Astrophysics} \textbf{\bibinfo{volume}{520}},
  \bibinfo{pages}{A28} (\bibinfo{year}{2010}).

\end{thebibliography}

\newcommand{\nop}[1]{}

\end{document}